\documentstyle[sprocl]{article}

\begin{document}

\title{RANDOMLY WALKING 1D QUANTUM HARMONIC OSCILLATOR.
AVERAGED TRANSITION PROBABILITIES.}

\author{A.S.GEVORKYAN, A.A.UDALOV}

\address{Institute for High-Performance Computing and Data Bases,\\
P/O Box 71, 194291, St-Petersburg, Russia\\
ashot@fn.csa.ru\quad  udalov@fn.csa.ru}

\maketitle\abstracts{One-dimensional problem for quantum harmonic
oscillator with
"regular+random" frequency subjected to the external "regular+random"
 force is considered. Averaged transition probabilities
are found.}
%\end{abstract}

\section{Introduction}

There are a lot of works published recently \cite{proc} and
devoted to quantum chaos, i.e. they consider quantum systems analogous to
classical systems that demonstrate the features of chaotic behaviour.
Exploration is developed at different directions:
investigation of energy levels distribution;
derivation and calculation of objects (analogous to
classical Lyapunov exponents or KS-entropy)
which can witness for chaos in quantum system and others.

We present a new approach for quantum description
of above mentioned systems that may be used for the investigation
of a problem of multichannel scattering. It is known
that scattering with rearrangement (typical for
chemical reactions) goes through resonance  complex formation
which causes chaotic behaviour. It is impossible to predict the
way to be followed by a system after leaving the complex
because of the small parameters change. As it was shown in our 
previous works,  \cite{bg1,bggm}
%^, \cite{bggm}
the scattering process with rearrangement may be described in
framework of randomly walking harmonic oscillator model.
In this paper we consider one-dimensional case.
More simple case has been already introduced earlier
\cite{bg2}.

\section{Description of the problem.}

We consider the wave function $ \Psi_{stc}(t,x), $
describing the state of the system as a random process
with the time evolution determined by equation
\begin{equation}
i\partial_{t}\Psi_{stc}=\hat H \Psi_{stc},
\label{001}
\end{equation}
where 1D Hamiltonian $ \hat H $ is quadratic in space
variable
\begin{equation}
\hat H = -\frac {1}{2}\frac {\partial^2}{\partial x^2}+
\frac {1}{2}\Omega^2(t)x^2-F(t)x,
\label{002}
\end{equation}
while functions $ \Omega^2(t) $  and  $ F(t) $
are random functions of time variable. We suppose that by definition
$$
\Omega^2(t) = \Omega_0^2(t) + \sqrt{2\epsilon_1p_1}f_1(t)\Theta(t-t_1),
$$
\begin{equation}
F(t) = F_0(t) + \sqrt{2\epsilon_2p_2}f_2(t)\Theta(t-t_2),
\label{003}
\end{equation}
where  $ \Omega_0^2(t) $  and  $ F_0(t)  $ are deterministic
functions while  $ f_1(t), f_2(t)  $
are independent zero mean gaussian random processes
with two-point correlations of $ \delta - $form:
\begin{equation}
<f_i(t)f_j(t')> = \delta_{ij}\delta (t-t'), \quad i,j = 1,2.
\label{004}
\end{equation}
Constants $ \epsilon_i,\ i=1,2 $ control the power of forces
$ f_i(t),\ t=1,2, $ while functions  $ p_i(t),\ i=1,2 $ are nonnegative
$ p_1,p_2\ge 0, $
which leads to the suggestion of the following asymptotic
behaviour
\begin{equation}
\Omega_0(t)\mathrel{\mathop{\longrightarrow }\limits_{t\rightarrow
\pm \infty }}\Omega_{\left (
in\atop out\right )},\quad
F_0(t)\mathrel{\mathop{\longrightarrow}\limits_{t\rightarrow
\pm \infty }} 0,\quad
p_i(t)\mathrel{\mathop{\longrightarrow}\limits_{t\rightarrow
\pm \infty }} 0,\ i=1,2.
\label{005}
\end{equation}
which guarantees in the limit  $ t\rightarrow -\infty $
existence of stationary states
$ \phi_n^{in}(t,x) $
\begin{eqnarray}
\phi_n^{in}(t,x) = e^{-i(n+1/2)\Omega_{in}t}\phi_n^{in}(x),
\quad \quad \quad \quad
\nonumber
\\
\phi_n^{in}(x) = \left (\frac {1}{2^nn!}\sqrt{\frac {\Omega_{in}}{\pi}}
\right )^{1/2}e^{-\Omega_{in}x^2/2}H_n(\sqrt{\Omega_{in}}x),
\label{006}
\end{eqnarray}
where $ H_n(x) $ are Hermitian polynomials.
At the limit $ t\rightarrow +\infty $
there also exist stationary states
$ \phi_n^{out}(t,x), $
which may be got from  (\ref{006})
by simple replacement of $ \Omega_{in} $
by $ \Omega_{out} $.
%$ \Theta $-functions in (\ref{003}) show that random
%processes $ f_1(t) $ and $ f_2(t) $ are switched on at
% $ t_1 $ and  $ t_2 $ respectively.
%If necessary $ p_1 $ and $ p_2 $ may be chosen in a way
%that ensures the absence of jumps in $ \Omega $ and $ F $
%at switching of the noise.
Moments  $ t_1 $ and $ t_2 $ of switching on the noise
are chosen to be finite in order to provide the correctness
of subsequent constructions.
Now we formulate our task: to get the averaged transition probabilities
$ W_{nm} $ from initial stationary states $ \phi_n^{in}(t,x) $
to final ones $ \phi_m^{out}(t,x) $ which come from the
evolution described by  (\ref{001})-(\ref{002}).

\section{Formal expressions for wave functional and
transition probabilities.}

The main results of this paper are found on the basis
of formal solution of (\ref{001})-(\ref{002}),
which may be constructed for arbitrary $ \Omega^2(t) $  and
$ F(t) $ as a functional of solutions of classical
equations of motion. Namely, as was shown in \cite{Baz'},
the following representation for solution of (\ref{001})-(\ref{002})
may be written that turns at the limit $ t\rightarrow -\infty $
to the stationary state $ \phi_n^{in}(t,x) $  defined in (\ref{006}):
\begin{equation}
\Psi_{stc}^{(n)}(t,x)=\frac {1}{\sqrt{r}}exp{\left \{i\left [\dot \eta (x-
\eta )+\frac {\dot r}{2r}(x-\eta )^2+\sigma \right ]\right \}}
\phi_n^{in}(\tau ,\frac {x-\eta }{r}),\quad n=1,2,...,
\label{101}
\end{equation}
where $ \eta (t) $ is a solution of the classical equation
of motion for oscillator with frequency $ \Omega(t) $
subjected to external force  $ F(t): $
\begin{equation}
\ddot \eta +\Omega^2(t)\eta =F(t),\quad \eta (-\infty )=
\dot \eta (-\infty )=0,
\label{102}
\end{equation}
$ \sigma (t) $ is an action functional corresponding to this
solution
\begin{equation}
\sigma (t)=\int_{-\infty }^t\left [\frac 1 2\dot \eta^2-
\frac 1 2\Omega^2\eta^2+F\eta \right ]dt',
\label{103}
\end{equation}
while $ r(t) $  and $ \tau (t) $ can be expressed in terms
of solution of homogeneous equation corresponding to (\ref{102})
\begin{equation}
\ddot \xi +\Omega^2(t)\xi =0,\quad
\xi (t)\mathrel{\mathop{\sim }\limits_{t\rightarrow -\infty }}
e^{i\Omega_{in}t}
\label{104}
\end{equation}
in the following way:
$\xi (t)=r(t)e^{\gamma (t)},\quad r(t)=\vert \xi (t)\vert,\quad
\tau =\gamma (t)/\Omega_{in}.$
Functions $ \Psi_{stc}^{(n)}(t,x) $ are in fact functionals
for the realisation of random processes  $ f_1(t) $ and $ f_2(t) $.
Therefore it is natural to call them "wave functionals".

We are interested in construction of transition probabilities
from the states $ \Psi_{stc}^{(n)}(t,x) $ to stationary states
$ \phi_m^{out}(t,x) $ in the limit $ t\rightarrow +\infty $
averaged over the random processes $ f_1(t) $ and $ f_2(t) $
realisations. Let's designate them as $ W_{nm}. $
Having defined
\begin{equation}
\Psi_{stc}^{(n)}(t,x)=\sum_{m=0}^\infty c_{nm}(t\vert f_1,f_2)
\phi_m^{out}(t,x),\quad
W_{nm} =\lim_{t\rightarrow +\infty }\left <\vert c_{nm}\vert^2
\right >,
\label{106}
\end{equation}
we denoted the averaging over $ f_1 $ and $ f_2 $ by the
symbol $ <..>. $ An expression for generating
function of coefficients $ c_{nm} $ is known (see \cite{Baz'}).
%$$
%I_{stc}(z_1,z_2,t)=\sum_{m,n=0}^\infty \frac {z_1^m}{\sqrt{m!}}
%c_{nm}(t\vert f_1,f_2)\frac {z_2^n}{\sqrt{n!}}.
%$$
%the following expression \cite{Baz'}:
%\begin{equation}
%I_{stc}(z_1,z_2,t)=\left (\frac {2\sqrt{\Omega_{in}\Omega_{out}}}{K\xi }
%\right )^{1/2}exp\{ Az_1^2+Bz_2^2+Cz_1z_2+Dz_1+Lz_2+M\}
%\label{107}
%\end{equation}
%where
%$$
%A=\frac 1 2e^{2i\Omega_{out}t}\left (\frac {2\Omega_{out}}{K}-1\right ),
%\quad B=\frac 1 2\left (\frac {2\Omega_{in}}{K\xi^2}-
%e^{-2i\gamma }\right ),
%$$
%$$
%C=\frac {2\sqrt{\Omega_{in}\Omega_{out}}}{K\xi }e^{i\Omega_{out}t}, \quad
%L=-\frac {\sqrt{2\Omega_{in}}}{K\xi }(\Omega_{out}\eta -i\dot \eta ),
%$$
%$$
%M=\frac {\Omega_{out}}{2}\left (\frac {\Omega_{out}}{K}-1\right )\eta^2
%-\frac {1}{2K}\dot \eta^2-\frac {i\Omega_{out}}{K}\eta \dot \eta +
%i\left ( \frac 1 2\Omega_{out}t+\sigma \right ),
%$$
%$$
%D=\sqrt{2\Omega_{out}}e^{i\Omega_{out}t}\left [\left (1-
%\frac {\Omega_{out}}{K}\right )\eta +\frac {i\dot \eta }{K}\right ],
%\quad K=-i\frac {\dot \xi }{\xi }+\Omega_{out }.
%$$
%Expanding (\ref{107}) into series of $ z_1 $ and  $ z_2 $
%powers one finds the coefficients $ c_{nm}. $
Having at hand formal expressions for objects of interest
we should turn the averaging procedure to the form
convenient for analytical or numerical treatment.

\section{Equation for the distribution function.}

Functions $ \xi (t) $  and  $ \eta (t) $ defined by equations
(\ref{102}), (\ref{104}) are random processes.
It is more convenient to treat with the other random processes
 $ z_1=\eta ,\ z_2=\dot \eta ,\
z_3=Re (\dot \xi /\xi ),\ z_4=Im (\dot \xi /\xi ) $
which in total we denote briefly as $ \vec z. $
%One obtains from (\ref{102}), (\ref{104}) the following set of equations
%for determination of components of vector random process $ \vec z $
%\begin{equation}
%\left\{
%\begin{array}{l}
%\dot z_1=z_2,\\
%\dot z_2=F_0-\Omega_0^2z_1-\sqrt{2\epsilon_1p_1}z_1f_1+
%\sqrt{2\epsilon_2p_2}f_2, \\
%\dot z_3=z_4^2-z_3^2-\Omega_0^2(t)-\sqrt{2\epsilon_1p_1}f_1(t),\\
%\dot z_4=-2z_3z_4,\qquad \qquad \qquad \qquad \qquad \qquad \qquad t>t_>,\\
%\end{array}\right.
%\label{402}
%\end{equation}
%$$
%\vec z(t_>)=\vec \zeta,
%$$
%where $ t_>\equiv {\rm max}(t_1,t_2), $ $ \ \vec \zeta $
%is an initial values vector which in general is also random.
%The set (\ref{402}) may be interpreted as a set of
%stochastic differential equations (SDE).
One can find that the Focker-Plank equation for
conditional distribution function
$ P(\vec z,t\vert \vec \zeta,t_>)=\left <\delta (\vec z(t)-
\vec z)\right >\left \vert_{
\vec z(t_>)=\vec \zeta
}\right. $
%corresponding to the set (\ref{402}) \cite{Gred,Gard}.
%^,\cite{Gard}
looks as follows:
\begin{equation}
\frac {\partial P}{\partial t}=-\sum_{i=1}^4
\frac {\partial (K_iP)}{\partial z_i}+(\epsilon_2p_2+
\epsilon_1p_1z_1^2)\frac {\partial^2 P}{\partial z_2^2}+
\epsilon_1p_1\frac {\partial^2 P}{\partial z_3^2}+
2\epsilon_1p_1z_1\frac {\partial^2 P}{\partial z_2
\partial z_3}\equiv \hat LP,
\label{403}
\end{equation}
where
$
K_1=z_2,\quad K_2=F_0-\Omega_0^2z_1,\quad K_3=z_4^2-z_3^2-
\Omega_0^2,\quad K_4=-2z_3z_4.
$
It is supplemented with the initial condition
$ P\vert_{t=t_>}=\delta (\vec z-\vec \zeta) $
and with the requirement that
$ \int P d\vec z $ is finite. Random vector $ \vec \zeta $ of initial
values of trajectories
is described by distribution function $ R(\vec \zeta,t_>) $ which
we do not specify without loss of information in final results.
%\begin{equation}
%R(\vec \zeta,t_>)=\left \{
%\begin{array}{l}
%P_1(\zeta_1,\zeta_2,t_>\vert z_{01},z_{02},t_<)\delta (\zeta_3-z_{03})
%\delta (\zeta_4-z_{04}),\qquad t_>=t_1, \\ \\
%P_2(\vec \zeta,t_>\vert \vec z_0,t_<), \qquad \qquad \qquad
%\qquad \qquad \qquad \qquad \qquad t_>=t_2, \\
%\end{array} \right.
%\label{401}
%\end{equation}
%where
%$ \vec z_0=\Bigl (\eta_0(t_>),
%\ \dot \eta_0(t_>),\ Re (\dot \xi_0(t_>)/\xi_0(t_>)),\ Im
%(\dot \xi_0(t_>)/\xi_0(t_>))\Bigr ), $
%$ t_<={\rm min}(t_1,t_2), $
%while $ P_1 $ and $ P_2 $ are determined from
%(\ref{403}) with $ \epsilon_1=0 $ and $ \epsilon_2=0 $
%respectively and initial conditions
%$$
%P_1\vert_{t=t_1}=\delta (z_1-z_{01})\delta (z_2-z_{02}),
%\quad P_2\vert_{t=t_2}=\delta (\vec z-\vec z_{0}).
%$$
%We have denoted as  $ \xi_0(t) $ and  $ \eta_0(t) $
%the solutions of regular equations corresponding to (\ref{102})
%and (\ref{104})
%\begin{equation}
%\ddot \xi_0 +\Omega_0^2(t)\xi_0 =0,\quad \xi_0(t)
%\mathrel{\mathop{\sim }\limits_{t\rightarrow -\infty }}
%e^{i\Omega_{in}t},
%\label{201}
%\end{equation}
%$$
%\xi_0(t)=r_0(t)e^{\gamma_0(t)}=\xi_{01}(t)+i\xi_{02}(t).
%$$
%\begin{equation}
%\ddot \eta_0 +\Omega_0^2(t)\eta_0 =F_0(t),\quad \eta_0(-\infty )=
%\dot \eta_0(-\infty )=0
%\label{202}
%\end{equation}

\section{Averaged transition probabilities.}

Averaged values of objects that are local in $ \vec z(t) $
are obtained by simple integration with weighting
function $ P(\vec z,t) $ from (\ref{403}).
Functionals in consideration $ W_{nm} $
are nonlocal in $ \vec z(t), $ but they have a special form
which allows to reduce the averaging procedure to solving some
parabolic differential equation \cite{4}. Namely, one can write down
the following
%formula \cite{4}:
%\begin{equation}
%\left <V_1(\vec z(t))\exp\left \{-\int_{t_>}^tV_2(\vec z(\tau ),
%\vec z(t),t)d\tau
% \right \} \right >=
%\int d\vec z V_1(\vec z)Q(\vec z,\vec z,t),
%\label{307}
%\end{equation}
%where the function $ Q(\vec z,\vec z',t) $ is a solution of the
%problem stated below
%\begin{eqnarray}
%\frac {\partial Q}{\partial t}=\left [\hat L(\vec z)-
%V_2(\vec z,\vec z ',t)\right ]Q,\qquad \quad
%\nonumber
%\\
%Q(\vec z,\vec z ',t)
%\mathrel{\mathop{\longrightarrow }\limits_{t\rightarrow t_1}}
%\delta (\vec z-\vec \zeta),
%\quad
% Q(\vec z,\vec z ',t)
%\mathrel{\mathop{\longrightarrow }\limits_{\vert \vert \vec u\vert
%\vert \rightarrow \infty }}0,
%\label{308}
%\end{eqnarray}
%where $ \vert \vert \cdot \vert \vert $ is some norm in $ R^4, $
%operator $ \hat L $ is defined in (\ref{403}).
%It is not difficult to deduce from (\ref{307}), (\ref{308})
%the following
representation for averaged transition probabilities
at arbitrary time $ t $
\begin{equation}
\left <\vert c_{nm}\vert^2\right >=
\int d\vec \zeta R(\vec \zeta,t_>)
\int d\vec z H_{nm}(\vec z)Q_{nm}(\vec z,t),
\label{404}
\end{equation}
%where against (\ref{307}) there is an extra integration
%with weighting function $ R(\vec \zeta,t_>) $ taking
%into account the averaging over the distribution
%of initial values of trajectories.
Functions $ H_{nm}(\vec z) $
are obtained from specific form of generating function for
coefficients $ c_{nm}. $ Functions $ Q_{nm}(\vec z,t) $
 include $ \vec \zeta $ as a parameter and are the solutions
of the following problem
\begin{eqnarray}
\frac {\partial Q_{nm}}{\partial t}=(\hat L-V_{nm})Q_{nm},
\quad \quad \qquad
\nonumber
\\
\quad Q_{nm}(\vec z,t)
\mathrel{\mathop{\longrightarrow }\limits_{t\rightarrow t_>}}
\delta (\vec z-\vec \zeta),
\quad
 Q_{nm}(\vec z,t)
\mathrel{\mathop{\longrightarrow }\limits_{\vert \vert \vec z\vert
\vert \rightarrow \infty }}0,
\label{405}
\end{eqnarray}
where
$ V_{nm}=p_{nm}z_3,\quad p_{00}=p_{01}=1,\quad p_{10}=p_{11}=3. $
Operator $ \hat L $ is defined in (\ref{403}).
Formulas (\ref{404})-(\ref{405}) are exact and give probabilities
$ W_{nm} $ at the limit $ t\rightarrow +\infty . $

Let's suppose that random forces $ f_1 $ and $ f_2 $
act with a constant power after switching on and then
are switched off at the moment $ t_e. $
%\begin{equation}
%p_i(t)=
%\left \{
%\begin{array}{l}
%1 \hspace {2cm} t_i<t<t_e,\\
%0 \hspace {2.4cm} t>t_e, \qquad i=1,2.\\
%\end{array}
%\right.
%\label{312}
%\end{equation}
Let's also assume that $ t_e $ is large enough to allow
the replacement of $ Q_{nm}(\vec z ,t_e) $ by the stationary
limit
$ Q_{nm}^{st}(\vec z)\equiv
\mathrel{\mathop{\lim }\limits_{t\rightarrow +\infty }}
Q_{nm}(\vec z,t). $
In such case one can obtain the
following expression for averaged probabilities
$ W_{nm} $
\begin{equation}
W_{nm}=\Omega_{in}^{p_{nm}}
\int d\xi_1d\xi_2d\xi_3
{\bar Q_{nm}^{st}(\xi_1,\xi_2,\xi_3)}
\bar H_{nm}\left (\xi_1, \xi_2, \xi_3\right ),
\label{326}
\end{equation}
where function $ \bar Q_{nm}^{st}(z_1,z_2,z_3) $
satisfies the stationary problem
%\begin{eqnarray}
$$
\left (-z_2\frac {\partial }{\partial z_1}+
\Omega_{out}^2z_1\frac {\partial }{\partial z_2}+
(z_3^2+\Omega_{out}^2)\frac {\partial }{\partial z_3}+
(\epsilon_2+\epsilon_1z_1^2)\frac {\partial^2 }{\partial z_2^2}+
\epsilon_1\frac {\partial^2 }{\partial z_3^2}+\right.
$$
%\qquad \qquad \qquad \qquad
%\nonumber
%\\
\begin{equation}
\left.+2\epsilon_1z_1\frac {\partial^2 }{\partial z_2
\partial z_3}\right )\bar Q_{nm}^{st}(\vec z )+
(2-p_{nm})z_3\bar Q_{nm}^{st}(\vec z)=0.
%\qquad \qquad\qquad \qquad \qquad \qquad \qquad
\label{322}
\end{equation}
and there are representations for the first
several functions $  \bar H_{nm}: $
%\begin{eqnarray}
$$
\bar H_{00}(\xi_1, \xi_2, \xi_3)=\frac {2\sqrt {\Omega_{in}\Omega_{out}}}
{\vert \xi_0(t_1)\vert \sqrt {\Sigma (\xi_3)}}
\exp \left \{-\frac {\Omega_{out}\Omega_{in}^2}{\Sigma (\xi_3)}
[\xi_3(\xi_1+\mu_1)-\xi_2-\mu_2]^2\right \}, \qquad
$$
%\nonumber
%\\
$$
\bar H_{01}(\xi_1, \xi_2, \xi_3)=
\frac {2\Omega_{out}\Omega_{in}^2}{\Sigma (\xi_3)}
[\xi_2-\xi_1\xi_3-\xi_3\mu_1+\mu_2]^2\bar H_{00}(\xi_1, \xi_2, \xi_3),
$$
%\qquad \qquad
%\nonumber
%\\
$$
\mu_1=-d_5+\sqrt {\frac {2\nu }{\Omega_{in}}}(d_1\cos \beta +d_2\sin \beta )
$$
%\qquad \qquad \qquad \qquad \qquad \qquad
%\nonumber
%\\
$$
\mu_2=-d_6+\sqrt {\frac {2\nu }{\Omega_{in}}}(d_3\cos \beta +d_4\sin \beta )
$$
%\qquad \qquad \qquad \qquad \qquad \qquad
%\nonumber
%\\
$$
\Sigma (\xi_3)=\frac {\Omega_{in}\Omega_{out}}{(1-\rho )}
\left \{ \biggl [(d_1^2+d_2^2)(1+\rho )-2\sqrt {\rho }
\Bigl [(d_1^2-d_2^2)\cos \delta +2d_1d_2\sin
\delta \Bigr ]\biggr ]\xi_3^2 \right.+
$$
%\nonumber
%\\
$$
\left.+2\biggl [-(d_2d_4+d_1d_3)(1+\rho )+2\sqrt {\rho }\Bigl [
(d_1d_4+d_2d_3)\sin \delta +
(d_1d_3-d_2d_4)\cos \delta \Bigr ]\biggr ]\xi_3+\right.
$$
%\nonumber
%\\
\begin{equation}
\left.+\biggl [(d_3^2+d_4^2)(1+\rho )+2\sqrt {\rho }
\Bigl [(d_4^2-d_3^2)\cos \delta -2d_3d_4\sin
\delta \Bigr ]\biggr ] \right \},
\label{337}
\end{equation}
where $ \delta =\delta_1+\delta_2 . $
For regular functions $ \xi_0(t) $ and $ \eta_0(t), $
defined by equations
$$
\ddot \xi_0 +\Omega_0^2(t)\xi_0 =0,\quad \xi_0(t)
\mathrel{\mathop{\sim }\limits_{t\rightarrow -\infty }}
e^{i\Omega_{in}t},
\
\xi_0(t)=\xi_{01}(t)+i\xi_{02}(t).
$$
$$
\ddot \eta_0 +\Omega_0^2(t)\eta_0 =F_0(t),\quad \eta_0(-\infty )=
\dot \eta_0(-\infty )=0
$$
we have used the following representations
%for asymptotic (at $ t\rightarrow \infty ) $ form
%of solution of (\ref{201})
$$
\xi_0(t)\mathrel{\mathop{\sim }\limits_{t\rightarrow +\infty }}
C_1e^{i\Omega_{out}t}+C_2e^{-i\Omega_{out}t},\quad
C_1=\vert C_1\vert e^{i\delta_1},\ C_2=\vert C_2\vert e^{i\delta_2},
$$
%for solution of (\ref{202})
$$
\eta_0(t)=\frac 1 {\sqrt{2\Omega_{in}}}\left (\xi_0d^*+\xi_0^*d\right ),
\quad d(t)=\frac i {\sqrt{2\Omega_{in}}}
\stackrel{t}{\mathrel{\mathop{\int }\limits_{-\infty }}}
\xi_0(t')F_0(t')dt',
$$
%and for concomitant quantities
Also we have used designations
$$
\rho =\left \vert \frac {C_2}{C_1}\right \vert^2,\quad
d=\lim_{t\rightarrow +\infty }d(t)=\sqrt {\nu }e^{i\beta }.
$$
$$
d_1=\xi_{01}(t_e),\ d_2=\xi_{02}(t_e),\ d_3=\dot \xi_{01}(t_e),\
d_4=\dot \xi_{02}(t_e),
$$
$$
d_5=\eta_0(t_e),\ d_6=\dot \eta_0(t_e).
$$

Obtained formula (\ref{326}) gives an approximate
value for $ W_{nm}. $  It is necessary to emphasise that it is
true only for finite $ \epsilon_1, \epsilon_2 , $
while the reducing $ \epsilon_1 $ or $ \epsilon_2 $ leads to increasing
the time  $ (t_e-t_>) $ necessary for setting the stationary distribution.

\end{document}